\documentclass[prd,twocolumn,aps,floats,floatfix,showpacs,showkeys,amsmath,amssymb,nofootinbib]{revtex4}

\usepackage{dcolumn}
\usepackage{bm}
\usepackage{txfonts}
\usepackage{pxfonts}
\usepackage{graphicx}
\usepackage{epsfig}
\usepackage{psfrag}
\usepackage{times}

\begin{document}
\title{Dissipation and acoustic tunnelling about the sonic horizon 
of Bondi accretion} 

\author{Arnab K. Ray}
\email{arnab_kumar@daiict.ac.in}
\affiliation{Dhirubhai Ambani Institute of Information and
Communication Technology, Gandhinagar 382007, Gujarat, India}

\date{\today}

\begin{abstract}
Viscous dissipation, as a small perturbative effect about the Bondi 
flow, shrinks its sonic sphere. An Eulerian perturbation on the steady 
flow gives a wave equation and the corresponding dispersion relation. 
The perturbation is a high-frequency
travelling acoustic wave, in which small dissipation is taken 
iteratively. The wave, propagating radially outwards against the 
bulk inflow, is blocked just within the sonic horizon, where 
the amplitude of the wave diverges because of viscosity.
The blocked acoustic wave can still tunnel outward 
through the horizon with a viscosity-dependent 
decaying amplitude, scaled by the analogue Hawking 
temperature. The escape of acoustic waves (analogue Hawking 
phonons) through the sonic
horizon is compatible with the radial contraction of the sonic
sphere. 
\end{abstract}


\maketitle

\section{Introduction} 
\label{sec1}
\citet{bon52} 
accretion is a classic textbook example of a 
transonic flow in astrophysics~\citep{fkr02,cc07}. This 
compressible astrophysical flow is steady and spherically 
symmetric, with its fluid elements being driven radially inwards 
by the gravity 
of a centrally located accretor, 
which can be an ordinary star or a neutron star or a 
black hole~\citep{pso80}. Far away from the accretor, the 
conservative velocity field has a very low subsonic value 
(idealized to vanish at infinity), but it becomes 
highly supersonic as it approaches the accretor. With the flow 
thus being subsonic at the outer boundary and supersonic at the 
inner boundary, it crosses the sonic barrier  
at an intermediate radius~\citep{bon52,rb02}. The surface of 
this sonic barrier, where the flow becomes transonic, is spherical. 
Travelling acoustic waves in the supersonic region are completely 
trapped within this spherical sonic surface. 
As a result the region 
bounded by the surface can be viewed as a spherically 
symmetric acoustic black hole~\citep{monc80,wgu81,wgu95,vis98},
and the sonic surface itself 
becomes a sonic horizon -- the acoustic analogue of the event 
horizon of a general-relativistic black hole. 

The distinctive features of~\citet{bon52} accretion, namely, 
spherically symmetric, compressible, convergent and irrotational, 
suit it well as a three-dimensional transonic potential flow. 
Natural examples of such flows are otherwise uncommon. 
These reasons make the~\citet{bon52} inflow, apart from 
being a paradigm in studies of astrophysical accretion, a 
physical model of interest in analogue gravity as well, from 
multiple perspectives~\citep{monc80,das04,macmal08,sr14}.   
In this work we take up~\citet{bon52} accretion 
as an analogue-gravity problem. Our aim is 
to investigate how the transonic conditions at the sonic horizon 
of conservative spherically 
symmetric accretion are affected by small viscous dissipation.  
Particularly since we know that the merest presence of viscosity 
disrupts
the acoustic geometry of conserved flows~\citep{vis98}. 
Certain effects of viscosity  
in spherically symmetric accretion have been taken
up in previous studies~\citep{an67,ray03}, but these have not 
explored the impact of viscous dissipation on the 
black-hole-like properties of the sonic horizon of transonic 
accretion. This is our specific quest. 

In Sec.~\ref{sec2}, we set down the governing hydrodynamic 
equations of a spherically symmetric inflow,  
incorporating viscous dissipation in the equation of momentum 
balance. 
In Sec.~\ref{sec3}, we study the impact of a perturbatively 
small order of viscosity on the steady transonic conditions
and estimate how viscosity contracts the sonic sphere, 
which points to the shrinking of acoustic black 
holes~\citep{wgu81,tj91,wgu95}. 
In Sec.~\ref{sec4}, by using an Eulerian perturbation scheme, 
we derive a viscous wave equation 
and establish the metric of an acoustic black hole in the 
inviscid limit. The wave equation also gives a dispersion relation
and the group velocity of the waves, using which we 
argue that viscosity diffuses the sharp
condition of the inviscid sonic horizon. Consistent with the 
acoustic black hole, in Sec.~\ref{sec5} we show that acoustic 
waves of high-frequency, travelling outwards against the 
radial fluid inflow, are blocked inside the sonic surface,
accompanied by an unstable growth in the amplitude of the wave.  
Cauchy's residue analysis at the sonic horizon, carried
out in Sec.~\ref{sec6}, allows the blocked wave to tunnel outward
through the impenetrable sonic barrier, with the viscosity-dependent
decaying amplitude of the wave being scaled by the 
analogue Hawking temperature. The weak viscosity-induced 
tunnelling concurs with the shrinking of the 
spherical acoustic black hole by viscosity.  
The acoustic tunnelling demonstrated here is the fluid 
analogue of Hawking radiation as a tunnelling 
phenomenon in general-relativistic cases~\citep{pw2000,pm07}. 
Hawking radiation as the tunnelling of acoustic waves through 
a critical point in a fluid flow is known for 
laboratory fluids~\citep{vol06,bhat17}. 
Here we establish the same for astrophysical accretion. 

In summary, the theme of this study is the fluid analogue of
gravity. The convergent inflow of viscous spherically symmetric 
accretion is an astrophysical fluid system where the formulations
of analogue gravity have been applied. Viscosity
is taken as a perturbative effect about 
the inviscid~\citet{bon52} inflow, which is a fundamental model 
of astrophysical accretion. The main results 
of this theoretical study are that viscous 
dissipation contracts the spherical sonic horizon, destabilizes
its neighbourhood under high-frequency perturbations, but lets
acoustic waves tunnel through the barrier. 

\section{The hydrodynamic equations} 
\label{sec2}
The flow of a viscous compressible fluid (in our study, 
an astrophysical gas flow, to be precise) is governed 
by the equation of continuity~\citep{ll87}, given as
\begin{equation}
\label{dcon}
\frac{\partial \rho}{\partial t}+
{\mathbf \nabla} \cdot \left(\rho {\mathbf v}\right)=0,
\end{equation}
and by the Navier-Stokes equation~\citep{ll87}, which, for a 
flow driven radially by the gravity of an accretor of mass, $M$, 
is expressed in full as 
\begin{equation}
\label{dns} 
\frac{\partial{\mathbf v}}{\partial t}+
\left({\mathbf v} \cdot {\mathbf \nabla}\right){\mathbf v}
+\frac{{\mathbf \nabla}P}{\rho}+\frac{GM}{r^2}\hat{\mathbf r} 
= \frac{1}{\rho}\left[\eta_1 \nabla^2{\mathbf v}+ 
\left(\frac{\eta_1}{3}+\zeta_2\right){\mathbf \nabla}
\left({\mathbf \nabla} \cdot{\mathbf v}\right)\right]. 
\end{equation} 
In the foregoing equation, $\eta_1$ and $\zeta_2$ are the
first (shear) and second (bulk) coefficients of viscosity,
respectively, both with positive values of the same
order~\citep{ll87}.
Since the bulk flow here is vorticity-free, the left hand side
of the vector identity,
${\mathbf \nabla}\times ({\mathbf \nabla}\times {\mathbf v})=
{\mathbf \nabla}({\mathbf \nabla} \cdot {\mathbf v}) 
- \nabla^2{\mathbf v}$, vanishes to give
${\mathbf \nabla}({\mathbf \nabla} \cdot {\mathbf v})
=\nabla^2 {\mathbf v}$, which we use to simplify the 
viscosity-dependent terms in Eq.~(\ref{dns}). Further, since 
accretion is a compressible fluid flow, 
${\mathbf \nabla}\cdot {\mathbf v}\neq 0$,
as seen in Eq.~(\ref{dcon}). 
The pressure, $P$, in Eq.~(\ref{dns}) is prescribed in terms of
the polytropic process~\citep{sc39} as 
\begin{equation} 
\label{polytro}
P=K\rho^\gamma, 
\end{equation} 
in which $\gamma$, the polytropic
exponent, is restricted by isothermal
and adiabatic limits over the range, 
$1 \leq \gamma \leq c_{\mathrm P}/c_{\mathrm V}$,
with $c_{\mathrm P}$ and $c_{\mathrm V}$ being the two
coefficients of specific heat capacity of a gas~\citep{sc39}.
With $P \equiv P(\rho)$, both the density, 
$\rho$, and the flow velocity, $\mathbf v$, are 
mathematically closed in Eqs.~(\ref{dcon}) and~(\ref{dns}). 

The viscosity of the accreting gas, being the outcome of 
internal friction on the molecular scale, is small. Viscosity 
of this order suits our purpose, as we argue now. In fundamental 
models of analogue gravity, the invariance of the acoustic geometry
requires an inviscid potential flow~\citep{wgu81,vis98}. 
To see how the conditions of this conservative background
are violated, we introduce viscosity as a feeble perturbative 
presence about the inviscid~\citet{bon52} inflow. We also know 
that the acoustic structure is violated 
on the molecular scale~\citep{vis98}, where the mean free path 
of the molecules is comparable to the short wavelength of acoustic 
waves. The known physical means of dissipation about this 
scale is, of course, molecular viscosity. 
Therefore, from this physical standpoint alone, a small molecular
viscosity suffices in our quest.
Besides the physics itself, 
the mathematical justification for considering a small molecular
viscosity lies in our linearized 
and iterative approach to a second-order nonlinear hydrodynamical 
system
(as we shall demonstrate from Sec.~\ref{sec3} onwards). 

Inasmuch as the molecular viscosity of 
a gas depends on its pressure and temperature (hence, on 
its density as well), the spatial variation of the molecular
viscosity is less than that of the bulk velocity of 
a spherically symmetric inflow. This is certainly true for 
a transonic radial flow in an open astrophysical
system, where radiative processes facilitate efficient cooling
and maintain the system close to being isothermal. By this 
argument, we approximate $\eta_1$ and $\zeta_2$ as 
constants~\citep{bh98,ray03}. 

Heating of a fluid, undergoing transonic accretion in spherical 
symmetry, can occur due to a variety of physical causes. 
These are   
viscosity~\citep{an67,ray03}, thermal conductivity~\citep{an67}, 
turbulence~\citep{tn89,rb05} and radiative 
transfer~\citep{ntz91} in both Newtonian 
gravity~\citep{an67,ray03,rb05} and 
Schwarzschild geometry~\citep{tn89,ntz91}.   
A polytropic process, which is more general than the limiting case
of an adiabatic process, encompasses the physical reality of heating 
in the fluid~\citep{sc39}. For a value of $\gamma$, the polytropic 
function in Eq.~(\ref{polytro}) forms a one-parameter 
family of curves~\citep{sc39}. This parameter, given by $K$ in 
Eq.~(\ref{polytro}), is related to 
the ``polytropic temperature", which becomes equal to the actual 
temperature of the fluid in the isothermal limit~\citep{sc39}. 
Eliminating $P$ in Eq.~(\ref{polytro}) by using the equation of state, 
$P=\rho k_{\mathrm B}T/\mu m_{\mathrm H}$~\citep{fkr02}, 
and then taking the logarithmic derivative, give 
\begin{equation}
\label{kayvar} 
\frac{{\mathrm d}K}{K} = \frac{{\mathrm d}T}{T}
- (\gamma -1)\frac{{\mathrm d}\rho}{\rho}, 
\end{equation} 
which shows how $K$ varies with the temperature, $T$, and 
the density~\citep{sc39}. 
With small viscous dissipation and efficient 
radiative cooling of the open astrophysical flow, the thermodynamic 
condition of the accreting fluid will be nearly isothermal. In that 
case, $\gamma \longrightarrow 1$ and $T$ will be almost fixed, 
which, considered together in Eq.~(\ref{kayvar}), will physically 
justify approximating $K$ as constant.

Tailored according to spherical symmetry, 
with $v\equiv v(r,t)$ and $\rho \equiv \rho (r,t)$, 
Eqs.~(\ref{dcon}) and~(\ref{dns}) are now written as 
\begin{equation}
\label{rcon}
\frac{\partial \rho}{\partial t}+\frac{1}{r^2}
\frac{\partial}{\partial r}\left(\rho vr^2\right)=0
\end{equation}
and 
\begin{equation}
\label{rns}
\frac{\partial v}{\partial t}+v\frac{\partial v}{\partial r}
+\frac{1}{\rho}\frac{\partial P}{\partial r}+\frac{GM}{r^2}
= \frac{\eta}{\rho}
\frac{\partial}{\partial r}
\left[\frac{1}{r^2} \frac{\partial}{\partial r} 
\left(vr^2\right)\right], 
\end{equation}
respectively~\citep{ll87}. The two coefficients
of viscosity are added in Eq.~(\ref{rns}), to
give a total viscosity, $\eta =(4/3)\eta_1 + \zeta_2$.
With Eqs.~(\ref{rcon}) and~(\ref{rns}), the 
coupled dynamics of $\rho$ and $v$ in the viscous 
spherically symmetric inflow is formulated completely. 

\section{The steady sonic horizon} 
\label{sec3}
In the steady state, partial time derivatives vanish, 
i.e. $\partial/\partial t \equiv 0$, leaving behind only 
full spatial derivatives. As such, integrating  
Eq.~(\ref{rcon}) gives 
\begin{equation} 
\label{coninteg}
4\pi \rho vr^2 = -\dot{m},
\end{equation} 
in which $\dot{m}$, a constant of the motion, is the matter 
inflow rate. Its negative sign, due to $v<0$, indicates the 
inward flux of matter~\citep{fkr02}. Using the constraint 
in Eq.~(\ref{coninteg})
and substituting the spatial derivatives of $\rho$ 
with the spatial derivatives of the local speed of sound, 
$a=\sqrt{\partial P/\partial \rho} =
\sqrt{\gamma K \rho^{\gamma -1}}$, we 
reduce Eq.~(\ref{rns}) in its steady limit 
($\partial v/\partial t =0$) to 
\begin{equation}
\label{srns}
\frac{\mathrm d}{{\mathrm d}r}\left(\frac{v^2}{2}+
\frac{a^2}{\gamma -1}-\frac{GM}{r}\right)=
\frac{2\eta}{(\gamma -1)\rho} 
\frac{\mathrm d}{{\mathrm d}r}\left(-\frac{v}{a}
\frac{{\mathrm d}a}{{\mathrm d}r}\right). 
\end{equation}
The viscosity-dependent term on the right 
hand side of Eq.~(\ref{srns}) is positive~\citep{ray03}, whose
physical implication is that viscosity opposes the gravity-driven
infall in spherical symmetry~\citep{ray03}. In contrast, by the
way, viscosity in accretion discs effects the outward transport
of angular momentum, and aids the axially symmetric 
infall~\citep{bh98,fkr02}. After substituting $\rho$ with $a$ 
in Eq.~(\ref{coninteg}), we replace the spatial derivatives 
of $a$ in Eq.~(\ref{srns}) 
by the spatial derivatives of $v$. This gives us
\begin{equation}
\label{dynam} 
\left(v-\frac{a^2}{v}-\frac{2\eta}{\rho r}\right)
\frac{{\mathrm d}v}{{\mathrm d}r}=
\frac{2a^2}{r}-\frac{GM}{r^2}+\frac{\eta}{\rho}
\left(\frac{{\mathrm d}^2 v}{{\mathrm d}r^2}-\frac{2v}{r^2}\right).
\end{equation}

When $\eta =0$, 
\citet{bon52} accretion becomes transonic at the sonic 
horizon~\citep{fkr02,rb02}. A smooth passage of the inflow 
through the sonic horizon requires both the right hand and the 
left hand sides of Eq.~(\ref{dynam}) to vanish together, while 
${\mathrm d}v/{\mathrm d}r \neq 0$, as in a first-order dynamical
system~\citep{rb02}. The values of $r$ and $v$ at the sonic 
horizon are thus obtained from Eq.~(\ref{dynam}) as
\begin{equation} 
\label{vrcrit}
r_{\mathrm c}=\frac{GM}{2a_{\mathrm c}^2}, \qquad
v_{\mathrm c}^2=a_{\mathrm c}^2,  
\end{equation} 
with the subscript ``$\mathrm c$" indicating the critical 
values when $\eta =0$. 
We now introduce the polytropic 
index, $n=(\gamma -1)^{-1}$~\citep{sc39}. Using it, 
the integral of Eq.~(\ref{srns}), for $\eta =0$, follows as 
\begin{equation}
\label{integeuler} 
\frac{v^2}{2} +na^2 -\frac{GM}{r} = E, 
\end{equation} 
with $E$ being the Bernoulli constant. The boundary
conditions of the flow are $v \longrightarrow 0$ and
$a \longrightarrow a_\infty$ (an ambient limit of $a$),
for $r \longrightarrow \infty$. Applying both boundary 
conditions   
in Eq.~(\ref{integeuler}) gives $E=na_\infty^2$. Further,  
putting the critical values 
given by Eqs.~(\ref{vrcrit}) in Eq.~(\ref{integeuler}), 
gives us $a_{\mathrm c}^2=2na_\infty^2/(2n-3)$. This fixes 
$v_{\mathrm c}$ and $r_{\mathrm c}$ in terms of the 
outer boundary conditions. Since $a$ is related to $\rho$, 
we also get 
$\rho_{\mathrm c}=\rho_\infty[2n/(2n-3)]^n$, with
$\rho \longrightarrow \rho_\infty$ (the ambient limit 
of $\rho$) for $r \longrightarrow \infty$. 

Our discussion so far has dwelt on the transonic inflow solution 
for $\eta =0$. But this solution is just one special 
solution in the $r$--$v$ phase plane, which also contains 
multiple classes of other solutions, both inflows and 
outflows~\citep{fkr02,rb02}. These 
solutions are not constrained by the critical conditions 
of Eqs.~(\ref{vrcrit}). For inflows, in which we are interested,
a general solution of $v(r)$ results from combining 
Eqs.~(\ref{coninteg})~and~(\ref{integeuler}).  
Eliminating $\rho$ and $a$, this solution with only $v$ and $r$ is 
\begin{equation} 
\label{vfr}
\frac{v^2}{2} + \frac{\gamma K}{\gamma -1} 
\left(\frac{-\dot{m}}{4\pi vr^2}\right)^{\gamma -1} 
- \frac{GM}{r} = E. 
\end{equation} 
Solving for $v(r)$ in Eq.~(\ref{vfr}) requires the prescription 
of two independent boundary conditions to determine the two 
independent constants of the motion, $\dot{m}$ and $E$. This 
is generally true for all $v(r)$. However, the transonic inflow
comes up with its own set of unique conditions, given by the 
critical coordinates in Eqs.~(\ref{vrcrit}). In such a situation,
$\dot{m}$ and $E$ are no longer independent of each other. 
Through the critical condition they become 
related to each other by 
\begin{equation}
\label{emdote} 
\dot{m}=\frac{\pi (GM)^2}{(\gamma K)^n}
\left(\frac{2E}{2n-3}\right)^{n-1.5}.
\end{equation} 
It is now clear that we can realize the transonic inflow in two
equivalent ways. One way is to provide the values of $\dot{m}$
and $E$ required for transonicity in Eq.~(\ref{vfr}). The 
alternative way (which is also the usual way) is through the 
critical coordinates in Eqs.~(\ref{vrcrit})
and specifying either $\dot{m}$ or $E$ 
(but not both) from a boundary condition. With viscosity 
included, we cannot obtain one constant of the motion, 
because Eq.~(\ref{srns}), with $\eta \neq 0$, no longer stands
for a conserved physical condition. The steady continuity 
condition, however, continues to be conservative. Hence, we 
are now left with only one constant of the motion. 
One might then suggest that at least for the 
transonic flow the loss of the second constant of 
the motion can be compensated by the critical conditions.  
But that will be mathematically self-contradictory, because 
if we do not have two independent constants of the motion, 
then its equivalent condition of a set of critical 
coordinates and one independent constant of the motion does
not stand either. Therefore, we conclude that for a viscous 
inflow our inability to find a constant of the motion from 
Eq.~(\ref{srns}) translates equivalently to the mathematical 
impossibility of a precise determination of the critical 
coordinates of the inflow. Deficiency of exact information 
about the critical coordinates in viscous transonic accretion 
has indeed been noted in studies on accretion~\citep{an67}. 

We now look at how the transonic conditions of 
Eq.~(\ref{dynam}), established clearly in the inviscid limit, 
are affected by viscosity, when it is treated as a small perturbative 
effect about the inviscid state~\citep{ray03}.  
Molecular viscosity is physically sufficient for this purpose, 
and we scale it in terms of quantities that are determined
by the microscopic molecular properties of the fluid on the 
large scale of the accretion radius, 
$r_{\mathrm a} \sim 2GM/a_\infty^2$, where 
gravity has no noticeable effect~\citep{fkr02}. 
We write $\eta = \eta_\star \eta_{\mathrm s}$, in which 
$\eta_\star$ is dimensionless and 
$\eta_{\mathrm s} = a_\infty \rho_\infty r_{\mathrm a}$. 
The latter sets the scale of $\eta$, and the 
former determines its magnitude, with $\eta_\star \ll 1$ 
in our perturbative analysis. 

Non-zero viscosity, howsoever small, renders Eq.~(\ref{dynam})
a differential equation of the second-order for a compressible 
fluid. Without the analytical solution of such a mathematical 
problem, we lose the precise sonic conditions, 
as set down in Eqs.~(\ref{vrcrit}), which are derived from a 
viscosity-free first-order dynamical system~\citep{rb02}. 
Nevertheless, the viscous 
flow will still be transonic~\citep{an67},\footnote{ 
In the thin boundary layer of a liquid flow, the second-order 
derivative in the viscous terms has a serious physical impact, 
but in the bulk its effect is not quite as qualitatively significant.
Spherically symmetric accretion is an open gas inflow with no boundary 
layer, and as such, with weak molecular viscosity, the bulk flow 
passes through a critical zone and becomes transonic~\citep{an67},
much like the inviscid inflow.} 
and for small $\eta$, 
the essential transonic features can be captured by a 
viscosity-dependent perturbative expansion in the neighbourhood 
of the inviscid equilibrium point~\citep{an67}. 
In a similar approach, we expand the viscous effect
about the inviscid sonic conditions of~\citet{bon52} accretion, 
by iteratively applying the inviscid 
velocity profile to the $\eta$-dependent second-order derivative 
of $v$ in Eq.~(\ref{dynam}). 
Near the accretor, for a freely falling inviscid
inflow, $v \sim r^{-1/2}$,
and far away from the accretor, for a highly subsonic
inflow, $v \sim r^{-2}$~\citep{pso80}. Going by these 
power laws in the two extreme limits of the inflow, we 
prescribe $v \sim r^{-\xi}$ for the steady velocity, 
with $0.5 < \xi < 2$ in the vicinity of the sonic point. 
Iteratively bringing in the inviscid power law of $v$ to the 
second-order derivative of $v$ on the right hand side of 
Eq.~(\ref{dynam}), reduces the derivative to an algebraic
form, $v/r^2$.\footnote{The second-order spatial derivative of 
$v$ is similarly approximated in the shallow flow of a
hydraulic jump (see~\citep{sbr05,rb07} and references therein).} 
The resulting first-order system lets us see how 
the inviscid critical coordinates, $r_{\mathrm c}$ and 
$v_{\mathrm c}$ in Eqs.~(\ref{vrcrit}), shift, 
respectively, to viscosity-dependent values, 
$r_{{\mathrm c}\eta}$ and $v_{{\mathrm c}\eta}$. 
To that end, the right and left hand sides of Eq.~(\ref{dynam}) 
give us the approximate conditions, 
\begin{equation} 
\label{aproxcrit1} 
2a_{\mathrm c}^2 r_{{\mathrm c}\eta} -GM \simeq 
-\frac{\eta v_{\mathrm c}}{\rho_{\mathrm c}} 
\left(\xi^2 +\xi -2\right), 
\end{equation} 
and 
\begin{equation}
\label{aproxcrit2} 
v_{{\mathrm c}\eta}-\frac{a_{\mathrm c}^2}{v_{{\mathrm c}\eta}}
\simeq \frac{2 \eta}{\rho_{\mathrm c} r_{\mathrm c}},
\end{equation} 
respectively. In Eqs.~(\ref{aproxcrit1}) and~(\ref{aproxcrit2}), 
terms with $\eta$ carry the inviscid critical values of 
$v_{\mathrm c}$, $\rho_{\mathrm c}$ and $r_{\mathrm c}$,  
which is consistent with our iterative approach to estimate 
the viscosity-induced correction on $v_{\mathrm c}$ and 
$r_{\mathrm c}$. Further, $a(r)$ is a slowly-varying 
function compared to the rate at which $v$ varies with $r$. 
This is very true when the flow becomes transonic and it is also 
consistent with the dependence of $a$ on the temperature, 
$T$, which, for a polytropic gas flow~\citep{sc39}, is 
$a \propto \sqrt{T}$. Efficient radiative processes in the 
open astrophysical flow maintain near-isothermal conditions, 
and as such, $a$ does not vary overmuch. With these arguments, 
and with viscosity as a small effect, we, therefore, retain 
the inviscid critical value of $a_{\mathrm c}$ in 
Eqs.~(\ref{aproxcrit1}) and~(\ref{aproxcrit2}).  
Now we divide Eq.~(\ref{aproxcrit1}) throughout 
by $2a_{\mathrm c}^2$. 
Then in the resulting $\eta$-dependent term, noting that 
$v_{\mathrm c} =-a_{\mathrm c}$ in the inviscid background 
flow, we relate $\eta$, $\rho_{\mathrm c}$ 
and $a_{\mathrm c}$ to $\eta_\star$, $n$, $a_\infty$, 
$\rho_\infty$ and $r_{\mathrm a}$, as has been outlined following 
Eq.~(\ref{integeuler}). Next, we relate $r_{\mathrm a}$ to 
$r_{\mathrm c}$. The physical part played by viscosity in the 
spherically symmetric inflow is to resist gravity and diminish
its field of attraction~\citep{ray03}.
At or about the critical point, the polytropic pressure term 
in Eq.~(\ref{rns}) is balanced against gravity~\citep{ray03}.
With gravity weakened by viscosity, the flow becomes transonic 
deeper within the gravitational potential well~\citep{ray03},
compared to the inviscid sonic radius. 
According to this physical reasoning, 
$r_{{\mathrm c}\eta} < r_{\mathrm c}$, 
which is achieved with $\xi \lesssim 1$.
The viscous spherically symmetric inflow now becomes  
transonic through a diffused region in the neighbourhood 
of $r_{\mathrm c}$. For small viscosity this region has a
thin radial width. To an order-of-magnitude, 
the fractional width of the transonic region, with a correction 
in $\eta_\star$, is  
\begin{equation} 
\label{rcrit} 
\frac{r_{{\mathrm c}\eta}-r_{\mathrm c}}{r_{\mathrm c}} \sim 
- \eta_\star \left(\frac{2n-3}{2n}\right)^{n-0.5}. 
\end{equation} 
The negative sign on the right hand side of Eq.~(\ref{rcrit}) 
shows that viscosity shrinks the sonic sphere of the 
inviscid transonic inflow to a reduced radius, 
$r_{{\mathrm c}\eta}$, when $\eta \neq 0$.
The physical significance of this result is that 
it holds viscous dissipation responsible for the loss of 
the surface area of the spherical sonic horizon~\citep{wgu81,tj91}. 
In Sec.~\ref{sec6}, we connect this viscosity-induced contraction
of the acoustic black hole to a viscosity-induced 
tunnelling of phonons through the acoustic horizon. 

When the flow is in the transonic stage, the inflow velocity 
also undergoes a small correction in proportion to $\eta_\star$. 
We estimate this correction by solving the quadratic equation 
of $v_{{\mathrm c}\eta}$ in Eq.~(\ref{aproxcrit2}). 
Neglecting the term with $\eta^2$ (for small $\eta$) in the 
discriminant of the quadratic solution, we get 
\begin{equation} 
\label{intervcrit} 
v_{{\mathrm c}\eta} \simeq \pm a_{\mathrm c} 
+ \frac{\eta}{\rho_{\mathrm c} r_{\mathrm c}}. 
\end{equation} 
For inflows, 
$v_{{\mathrm c}\eta} \simeq -[a_{\mathrm c}-(\eta/
\rho_{\mathrm c} r_{\mathrm c})]$, and for outflows, 
$v_{{\mathrm c}\eta} \simeq a_{\mathrm c}+(\eta/
\rho_{\mathrm c} r_{\mathrm c})$. Evidently, $\eta$ breaks the 
symmetry between the two critical speeds. 
Pertaining to inflows, we
relate $\eta$ and $\rho_{\mathrm c}$ to ambient values 
at the outer boundary, as done to derive Eq.~(\ref{rcrit}).  
Then we arrive at the approximate result, 
\begin{equation}
\label{vcrit} 
v_{{\mathrm c}\eta} \simeq -a_{\mathrm c} +4a_{\mathrm c} 
\eta_\star \left(\frac{2n-3}{2n}\right)^{n-0.5}, 
\end{equation}
in which the smallness of $\eta_\star$ ensures that the 
Mach number is almost unity in the transonic region~\citep{an67}. 
From Eqs.~(\ref{rcrit}) and~(\ref{vcrit}) we get an approximate 
idea of the sonic conditions of the compressible viscous 
spherically symmetric inflow.

The sonic coordinates of the inviscid~\citet{bon52} solution 
are determined with full precision from the single equilibrium point 
of the first-order dynamical system in Eq.~(\ref{dynam}), 
when $\eta =0$~\citep{rb02}. This precision is lost as soon 
as viscosity is operative, although a transonic solution still 
exists~\citep{an67,tn89}. To go past this difficulty we 
have prescribed 
the flow velocity as a power law of the radial distance, 
and thus obtained Eq.~(\ref{aproxcrit1}). This approach is 
convenient when viscosity is in effect~\citep{ray03}, and its
quantitative results do not deviate significantly from the 
inviscid case, unless the effective viscosity is high,
such as in turbulence~\citep{tn89,rb05}.  
In fact, even the inviscid~\citet{bon52} accretion admits 
self-similar solutions~\citep{gai06}, established by power 
laws of the velocity and density profiles.
With self-similar profiles, it happens that
one power law is valid near the centre of coordinates
and another is valid asymptotically at the outer
boundary. We have seen this feature while arriving
at Eq.~(\ref{aproxcrit1}). The transition from one power law
to another takes place about the critical point. In many cases,
viscosity is necessary for the simultaneous existence of two
distinct self-similar solutions, each separated from the other 
by a narrow crossover zone. This is known well in a wide range 
of hydrodynamical problems, for example, gaseous baryonic structures
on cosmological scales~\citep{hs03}, and the shallow-water
hydraulic jump on laboratory scales~\citep{bdp93,sbr05,rb07}.
The inner and outer boundary conditions of these 
systems often select the physically relevant self-similar 
solutions, which in turn determine the position of the critical 
point~\citep{bdp93,sbr05,rb07}. Inappropriate 
choice of boundary conditions can, therefore, result in extraneous 
critical points. In the particular context of spherically symmetric 
accretion on to a Schwarzschild black hole, more than one 
critical point do appear~\citep{ntz91,mrd07,cms16}. 
However, we do not expect such a situation to come about
in our present case of a viscous inflow, since it involves only 
a small viscosity under  
Newtonian gravity. In fact, our results about the critical point, 
especially Eq.~(\ref{intervcrit}), are independently supported 
by the group velocity of radially propagating waves that we get
by perturbing the steady flow. We show this in Sec.~\ref{sec4}.   

\section{A wave equation} 
\label{sec4} 
In the steady state, the solutions of 
Eqs.~(\ref{rcon}) and~(\ref{rns}) are two coupled time-independent 
fields, that we write as $\rho_0(r)$ and $v_0(r)$. About these
steady flow profiles, we impose small time-dependent radial 
perturbations and then linearize the perturbed quantities.
The prescription for the time-dependent radial perturbation
is $v(r,t)=v_0(r)+v^{\prime}(r,t)$
and $\rho (r,t)=\rho_0(r)+\rho^{\prime}(r,t)$, with 
the primed quantities being perturbations about a
steady background. Following an Eulerian perturbation
scheme employed by~\citet{pso80}, we define a new variable, 
$f(r,t)=\rho vr^2$, which emerges
as a constant of the motion from the steady limit of
Eq.~(\ref{rcon}). This constant, $f_0=\rho_0 v_0r^2$, 
is the matter flow rate, within a geometrical
factor of $4\pi$~\citep{fkr02}, as seen in 
Eq.~(\ref{coninteg}). On applying the perturbation
scheme for $v$ and $\rho$, the perturbation in $f$
is derived as
\begin{equation}
\label{pertef}
\frac{f^\prime}{f_0} = \frac{\rho^\prime}{\rho_0} 
+\frac{v^\prime}{v_0}, 
\end{equation}
connecting $v^\prime$, $\rho^\prime$ and $f^\prime$. 
To relate $\rho^\prime$ and $f^\prime$ to each other, 
we apply the perturbation scheme to Eq.~(\ref{rcon}),
resulting in
\begin{equation}
\label{pertrho}
\frac{\partial \rho^\prime}{\partial t}
= -\frac{1}{r^2} 
\frac{\partial f^\prime}{\partial r}.
\end{equation}
To obtain a similar relation between $v^\prime$ and
$f^\prime$, we combine the conditions
given in Eqs.~(\ref{pertef}) and~(\ref{pertrho}), to get
\begin{equation}
\label{pertvee}
\frac{\partial v^\prime}{\partial t} = \frac{v_0}{f_0}
\left(\frac{\partial f^\prime}{\partial t}+ 
v_0 \frac{\partial f^\prime}{\partial r} \right).
\end{equation}
Together, Eqs.~(\ref{pertrho}) and~(\ref{pertvee}) form a
closed set, with $\rho^\prime$ and $v^\prime$ 
expressed exclusively in terms of $f^\prime$. We now 
need an independent condition, in which we can
use Eqs.~(\ref{pertrho}) and~(\ref{pertvee}). Just such a 
condition is offered by Eq.~(\ref{rns}). 
We take the second-order time derivative of Eq.~(\ref{rns}), 
to which we then apply Eqs.~(\ref{pertrho}),~(\ref{pertvee}) 
and the second-order time derivative of Eq.~(\ref{pertvee}). 
This exercise leads us to the wave equation,
\begin{multline}
\label{perteq}
\frac{\partial}{\partial t}
\left(h^{tt}\frac{\partial f^\prime}{\partial t}\right) +
\frac{\partial}{\partial t}
\left(h^{tr}\frac{\partial f^\prime}{\partial r}\right) +
\frac{\partial}{\partial r}
\left(h^{rt}\frac{\partial f^\prime}{\partial t}\right)  
+ \frac{\partial}{\partial r}
\left(h^{rr}\frac{\partial f^\prime}{\partial r}\right) \\ 
= \eta v_0r^2
\frac{\partial}{\partial r}\left[\frac{1}{r^2}
\frac{\partial}{\partial r}\left(\frac{1}{\rho_0}
\frac{\partial f^\prime}{\partial t}+\frac{v_0}{\rho_0}
\frac{\partial f^\prime}{\partial r}\right)\right] 
+\frac{\eta v_0}{\rho_0}\frac{\partial f^\prime}{\partial r} \times \\
\frac{\partial}{\partial r}\left[\frac{1}{r^2}
\frac{\partial}{\partial r}\left(v_0 r^2\right)\right],
\end{multline}
in which $h^{tt}=v_0$, $h^{tr}=h^{rt}=v_0^2$ and 
$h^{rr}=v_0(v_0^2 -a_0^2)$, with $a_0$ in $h^{rr}$ being the 
steady value of $a$. 

We find it instructive to examine Eq.~(\ref{perteq}) 
for $\eta=0$. In this inviscid limit, 
going by the symmetry of the left hand side of Eq.~(\ref{perteq}), 
we can recast it in a compact form as 
$\partial_\mu \left(h^{\mu \nu}\partial_\nu f^\prime \right) =0$,
with the Greek indices running from $0$ to $1$, under the
equivalence that $0$ stands for $t$ and $1$ stands for $r$. 
All the $h^{\mu \nu}$ in Eq.~(\ref{perteq}) are to be seen as 
elements of the matrix,
\begin{equation}
\label{matrix}
h^{\mu \nu }=v_0
\begin{pmatrix}
1 \hfill & v_0 \\
v_0  & v_0^2 - a_0^2 \hfill
\end{pmatrix}. 
\end{equation}
Now, in Lorentzian geometry the d'Alembertian of a scalar field
in curved space is obtained from the metric, $g_{\mu \nu}$, as
\begin{equation}
\label{alem}
\Delta \varphi \equiv \frac{1}{\sqrt{-g}}
\partial_\mu \left({\sqrt{-g}}\, g^{\mu \nu}
\partial_\nu \varphi \right),
\end{equation}
where $g^{\mu \nu}$ is the inverse of the matrix,
$g_{\mu \nu}$~\citep{monc80,vis98,blv11}. Comparing
Eq.~(\ref{alem}) with Eq.~(\ref{perteq}) (when $\eta =0$), we 
identify $h^{\mu \nu }=\sqrt{-g}\, g^{\mu \nu}$~\citep{vis98}, 
and see that the wave equation of $f^\prime$ in Eq.~(\ref{perteq}) 
is similar to Eq.~(\ref{alem}). 
The metric that is implicit in Eq.~(\ref{perteq}), is to be 
read from Eq.~(\ref{matrix}), 
and its inverse establishes an acoustic metric and an 
acoustic horizon, when $v_0^2 = a_0^2$~\citep{vis98}. 
In the radial inflow of the~\citet{bon52} accretion, this 
horizon is due to an acoustic black hole.
The radius of the horizon is the critical radius, $r_{\mathrm c}$,
given in Eqs.~(\ref{vrcrit}), which 
cannot be breached by any acoustic wave 
propagating against the bulk inflow,
after having originated in the supercritical region,
where $v_0^2 > a_0^2$ and $r < r_{\mathrm c}$.
Borne by the wave, the flow of a signal across 
the acoustic horizon is, therefore, only inwards. 
With the inclusion of viscosity ($\eta \neq 0$), however, a source 
term appears in Eq.~(\ref{perteq}) to compromise the symmetry 
of the acoustic metric, which we can relate to the violation 
of the acoustic Lorentz invariance by viscous dissipation~\citep{vis98}. 
Without viscosity the acoustic horizon is determined precisely when  
$v_0 = \pm a_0$, the positive sign of $a_0$ standing for outflows and 
the negative sign for inflows. Accounting for viscosity, 
perturbatively small albeit, the acoustic horizon is 
diffused about the sharp inviscid condition, $v_0 = \pm a_0$. 
For small viscosity, this loss of precision has been approximately
estimated in Eq.~(\ref{intervcrit}). 

This point of view is also favoured by the dispersion relation
that we can extract from Eq.~(\ref{perteq}), and the consequent 
group velocity of the acoustic waves. With respect to a 
homogeneous fluid background,  
a viscous wave equation is yielded by Eq.~(\ref{perteq}) as, 
\begin{equation} 
\label{visweq} 
\frac{\partial^2 f^\prime}{\partial t^2}=a_0^2 
\frac{\partial^2 f^\prime}{\partial r^2}+
\frac{\eta}{\rho_0} \frac{\partial}{\partial t} 
\left(\frac{\partial^2 f^\prime}{\partial r^2}+
\frac{2}{r} \frac{\partial f^\prime}{\partial r}-
\frac{2f^\prime}{r^2}\right), 
\end{equation} 
in a familiar form~\citep{vis98}. The solution,
$f^\prime (r,t) \sim \exp [i(kr-\omega t)]$, 
applied to Eq.~(\ref{visweq}), gives a quadratic equation,
\begin{equation} 
\label{dispquad} 
\left(\omega -kv_{\mathrm B}\right)^2 -\frac{i\eta}{\rho_0} 
\left(-k^2 -\frac{2}{r^2}+\frac{2ik}{r}\right)
\left(\omega -kv_{\mathrm B}\right) -k^2a_0^2 =0, 
\end{equation} 
in which $v_{\mathrm B}$ stands for the bulk 
motion of the fluid. For small viscosity, we neglect terms with 
$\eta^2$ in the discriminant of the quadratic form in 
Eq.~(\ref{dispquad}), to get a dispersion relation, 
\begin{equation} 
\label{disper} 
\omega \simeq kv_{\mathrm B} \pm ka_0 - \frac{\eta k}{\rho_0 r} 
- \frac{i\eta}{2\rho_0} \left(k^2 + \frac{2}{r^2}\right). 
\end{equation}   
Dissipation happens because of the viscosity-dependent imaginary 
term in Eq.~(\ref{disper}). The real terms, taken together, 
contribute to the group velocity of the wave, 
\begin{equation} 
\label{groupvel} 
v_{\mathrm g} = \frac{\partial \omega}{\partial k} \simeq 
v_{\mathrm B} \pm a_0 - \frac{\eta}{\rho_0 r}. 
\end{equation} 
First of all, the wave is carried by the bulk motion of the fluid, 
with a velocity, $v_{\mathrm B}$. With respect to the fluid itself, 
outgoing waves have the group velocity, $a_0 -(\eta/\rho_0 r)$, and 
incoming waves have the group velocity, $-[a_0 +(\eta/\rho_0 r)]$. But 
for $\eta \neq 0$, the two waves would have had the same speed, 
and the sonic points of inflows and outflows would have coincided.
The breaking of symmetry in the group velocities of outgoing and 
incoming waves, entirely due to viscosity, agrees with what 
Eq.~(\ref{intervcrit}) suggests -- that the acoustic horizons 
of inflows and outflows of a viscous compressible fluid are 
diffused about the inviscid sonic condition, 
$v_{\mathrm c} = \pm a_{\mathrm c}$, by a narrow velocity width 
of approximate order, $\eta/(\rho_{\mathrm c}r_{\mathrm c})$. 
The viscous inflow becomes transonic through this non-zero velocity 
width, and not through a sharply defined sonic point, as in 
Eqs.~(\ref{vrcrit}).  

\section{Wave blocking and instability about the sonic horizon} 
\label{sec5}
Without viscosity on the right hand side of 
Eq.~(\ref{perteq}), linear perturbations do not destabilize 
the steady background flow~\citep{pso80}. We now look at 
the effect that viscosity has on this condition. 
We treat the perturbation as a 
high-frequency travelling wave, whose wavelength, $\lambda$, 
is much less than the natural length scale in the inflow, 
the radius of the acoustic horizon, $r_{\mathrm c}$. Thus,  
specifying $\lambda \ll r_{\mathrm c}$, we use a separable 
solution for the travelling wave as 
$f^\prime (r,t) =\exp [is(r)-i\omega t]$,
with the understanding
that $\omega$ is much greater than any characteristic frequency
of the system. Applying the foregoing solution to Eq.~(\ref{perteq})
and carrying out algebraic simplifications, deliver
\begin{multline} 
\label{sepsol}
\left(v_0^2 - a_0^2 \right)
\left[i\frac{{\mathrm d}^2 s}{{\mathrm d}r^2}
-\left(\frac{{\mathrm d}s}{{\mathrm d}r}\right)^2\right]
+\frac{i}{v_0}
\frac{{\mathrm d}}{{\mathrm d}r}
\left[v_0 \left(v_0^2 -a_0^2\right)\right]
\frac{{\mathrm d}s}{{\mathrm d}r} \\
+2v_0\omega \frac{{\mathrm d}s}{{\mathrm d}r}
-2i\omega \frac{{\mathrm d}v_0}{{\mathrm d}r}
-\omega^2 =\eta \Bigg\{-\frac{iv_0}{\rho_0}
\left(\frac{{\mathrm d}s}{{\mathrm d}r}\right)^3 \\
-\frac{3v_0}{\rho_0}\frac{{\mathrm d}s}{{\mathrm d}r}
\frac{{\mathrm d}^2s}{{\mathrm d}r^2} 
+\left(\frac{{\mathrm d}s}{{\mathrm d}r}\right)^2
\left[\frac{2v_0}{r\rho_0}
-2\frac{\mathrm d}{{\mathrm d}r}\left(\frac{v_0}{\rho_0}\right)  
+\frac{i\omega}{\rho_0}\right]
+\frac{iv_0}{\rho_0}\frac{{\mathrm d}^3s}{{\mathrm d}r^3} \\
+\frac{{\mathrm d}^2s}{{\mathrm d}r^2} 
\bigg[\frac{\omega}{\rho_0}-\frac{2iv_0}{r\rho_0}-2i
\frac{\mathrm d}{{\mathrm d}r}\left(\frac{v_0}{\rho_0}\right)\bigg] 
+\frac{{\mathrm d}s}{{\mathrm d}r}\bigg[i
\frac{{\mathrm d}^2}{{\mathrm d}r^2}\left(\frac{v_0}{\rho_0}\right)
-\frac{2i}{r}\frac{\mathrm d}{{\mathrm d}r}\left(\frac{v_0}{\rho_0}\right) \\
-\frac{i}{\rho_0} 
\frac{\mathrm d}{{\mathrm d}r}\left(\frac{v_0}{\rho_0}
\frac{{\mathrm d}\rho_0}{{\mathrm d}r}\right)
-\frac{2\omega}{r\rho_0}-\frac{2\omega}{\rho_0^2} 
\frac{{\mathrm d}\rho_0}{{\mathrm d}r}\bigg]
-\frac{2i\omega}{r\rho_0^2}\frac{{\mathrm d}\rho_0}{{\mathrm d}r}
-i\omega\frac{{\mathrm d}^2\rho_0^{-1}}{{\mathrm d}r^2}
\Bigg\}.
\end{multline} 
From Eq.~(\ref{sepsol}) it is clear that $s$ has both real 
and imaginary parts. Therefore, we prescribe $s(r)=\alpha (r)+i\beta (r)$,
with both $\alpha$ and $\beta$ being real.
Going by the separable form of $f^\prime$, 
we see that while $\alpha$ contributes to the phase of the 
perturbation, $\beta$ contributes to its amplitude. 
Solutions of both $\alpha$ and $\beta$ are found by a 
{\it WKB} analysis of Eq.~(\ref{sepsol}), according to which we 
consider $\alpha \gg \beta$ for travelling waves of high frequency.
In Eq.~(\ref{sepsol}), the term of the highest order in $s$ 
is cubic. Fortunately, this term is 
dependent on $\eta$, whose presence in our analysis is very feeble 
anyway. We use this fact to our advantage by first setting $\eta =0$ 
in Eq.~(\ref{sepsol}), after which we solve a  
second-order differential equation that is independent of $\eta$. 
In this special case of $\eta =0$, we also modify our solution 
as $s_0(r)= \alpha_0 (r)+i\beta_0 (r)$, with the subscript ``$0$"
denoting solutions in the absence of viscosity.
Using this in the inviscid limit of Eq.~(\ref{sepsol}), we
first separate the real and the imaginary parts, which are then
individually set equal to zero. The {\it WKB} prescription stipulates
that $\alpha_0 \gg \beta_0$. Accordingly we 
collect only real terms without $\beta_0$, and solve
a resulting quadratic equation
in ${\mathrm d}\alpha_0/{\mathrm d}r$ to obtain
\begin{equation}
\label{alphanot}
\alpha_0 = \int \frac{\omega}{{v_0 \mp a_0}}\,{\mathrm d}r.
\end{equation}
Likewise, from the imaginary part, in which we need to use the
solution of $\alpha_0$, we obtain
\begin{equation}
\label{betanot}
\beta_0 = \frac{1}{2} 
\ln \left(v_0 a_0\right)+ C,
\end{equation} 
with $C$ being a constant of integration~\citep{sbbr13}. 

We now perform a self-consistency check that $\alpha_0 \gg \beta_0$,
as a basic requirement of our {\it WKB} analysis. First, we note 
$\alpha_0$ contains $\omega$ (the high frequency of the travelling 
wave), and in this respect is of a 
leading order over $\beta_0$, which contains $\omega^0$. Next,
on very large scales of length, i.e. $r \longrightarrow \infty$,
the asymptotic behaviour of the background velocity is
$v_0 \longrightarrow 0$, and the corresponding speed of acoustic
propagation, $a_0$, approaches a
constant asymptotic value. In that case, $\alpha_0 \sim \omega r$
in Eq.~(\ref{alphanot}). Moreover, on similar scales of
length, going by $v_0 \sim f_0 r^{-2}$, 
we see that $\beta_0 \sim \ln r$. Further, near the acoustic horizon,
where $\vert v_0 \vert \simeq a_0$, for the wave that goes against 
the bulk inflow with the speed, $a_0 -\vert v_0 \vert$, there is 
a singularity in $\alpha_0$. All of these facts taken together, we
see that our solution scheme is well in conformity with the
{\it WKB} prescription.

Thus far we have worked with $\eta =0$ (absence of viscosity).
To know how viscous dissipation affects the travelling wave, we 
have to find a solution of $s$ from Eq.~(\ref{sepsol}), with 
$\eta \neq 0$. To this end, we adopt an iterative approach, 
exploiting the condition that $\eta = \eta_\star \eta_{\mathrm s}$ 
has a very small value, the smallness being set by $\eta_\star \ll 1$. 
Then taking up Eq.~(\ref{sepsol}) in full, we  
propose a solution for it as $s=s_0 + \delta s_\eta$, with
$\delta$ being another dimensionless parameter that, like $\eta_\star$,
obeys the requirement, $\delta \ll 1$.
Therefore, on the right hand side of Eq.~(\ref{sepsol}) all terms
that carry the product, $\eta_\star \delta$, can be safely neglected
as being very small. This, in keeping with the principle of our
iterative treatment, effectively means that all the
surviving viscosity-related terms
on the right hand side of Eq.~(\ref{sepsol})
will go as $\eta s_0$. Further, by the {\it WKB} analysis,
we have also assured ourselves that $\alpha_0 \gg \beta_0$, by 
which we ignore all dependence on $\beta_0$ 
on the right hand side of Eq.~(\ref{sepsol}), when we compare
them with all the terms containing $\alpha_0$. With these
arguments, we approximate $s \simeq \alpha_0$ on the right hand 
side of Eq.~(\ref{sepsol}), and see here that 
the most dominant $\alpha_0$-dependent real terms
are of the quadratic degree.
Preserving only these terms on the right hand side of Eq.~(\ref{sepsol})
and extracting only the $\beta$-independent real terms from the
left hand side, we arrive at a quadratic equation in 
${\mathrm d}\alpha/{\mathrm d}r$,
\begin{multline}
\label{alphaiter}
\left(v_0^2 -a_0^2 \right)
\left(\frac{{\mathrm d}\alpha}{{\mathrm d}r}\right)^2
-2v_0 \omega
\frac{{\mathrm d}\alpha}{{\mathrm d}r}
+ \omega^2 +\eta \Bigg\{
-\frac{3v_0}{\rho_0}\frac{{\mathrm d}\alpha_0}{{\mathrm d}r}
\frac{{\mathrm d}^2\alpha_0}{{\mathrm d}r^2} \\
+2\left(\frac{{\mathrm d}\alpha_0}{{\mathrm d}r}\right)^2
\left[\frac{v_0}{r\rho_0}
-\frac{\mathrm d}{{\mathrm d}r}\left(\frac{v_0}{\rho_0}\right)  
\right]\Bigg\} =0.
\end{multline}
We solve Eq.~(\ref{alphaiter}) under the provision of
$\eta_\star \ll (\lambda/r_{\mathrm c})^2$, which accords 
well with our requirement that $\eta_\star$ may be arbitrarily
small. A binomial approximation of 
terms with $\eta$ in the discriminant gives us 
$\alpha= \alpha_0 + \alpha_\eta$, with the viscosity-dependent 
correction to $\alpha$ being 
\begin{equation} 
\label{alphacor}
\alpha_\eta \simeq \pm 
\int 
\frac{\eta \omega v_0}{2r\rho_0 a_0\left(v_0 \mp a_0\right)^2}
\bigg[2+\frac{4r}{v_0}\frac{{\mathrm d}v_0}{{\mathrm d}r} 
-\frac{3}{v_0\mp a_0}\frac{\mathrm d}{{\mathrm d}r} 
\left(v_0\mp a_0\right)\bigg]{\mathrm d}r.
\end{equation} 

Next, to take up $\beta$, we extract all the imaginary
terms from the left hand side of Eq.~(\ref{sepsol}), and noting
that the most dominant contribution to the imaginary terms on
the right hand side is of the cubic degree in 
$\alpha_0$, we need to solve the equation,
\begin{multline}
\label{betaiter}
2\left[v_0 \omega -\left(v_0^2 -a_0^2\right)
\frac{{\mathrm d}\alpha}{{\mathrm d}r}\right]
\frac{{\mathrm d}\beta}{{\mathrm d}r} 
+\frac{1}{v_0}
\frac{\mathrm d}{{\mathrm d}r}
\left[v_0 \left(v_0^2 -a_0^2\right)
\frac{{\mathrm d}\alpha}{{\mathrm d}r}\right] \\
-2\omega\frac{{\mathrm d}v_0}{{\mathrm d}r} 
= -\frac{\eta v_0}{\rho_0}
\left(\frac{{\mathrm d}\alpha_0}{{\mathrm d}r}\right)^3.
\end{multline}
We observe that $\alpha_0$ and $\alpha_\eta$ are both 
linear in $\omega$, whereas the right hand 
side of Eq.~(\ref{betaiter}), with 
$({\mathrm d}\alpha_0/{\mathrm d}r)^3$, is cubic in  
$\omega$. Therefore, the dominant correction in $\beta$ 
due to viscosity can be found with the approximation 
$\alpha \simeq \alpha_0$ on the left hand side of 
Eq.~(\ref{betaiter}). This reasoning gives us 
$\beta = \beta_0 +\beta_\eta$, in which 
\begin{equation}
\label{betacor} 
\beta_\eta \simeq \pm 
\int 
\frac{\eta \omega^2 v_0}{2\rho_0 a_0\left(v_0 \mp a_0\right)^3}\,
{\mathrm d}r.
\end{equation} 
A noteworthy aspect of both
Eqs.~(\ref{alphacor}) and~(\ref{betacor}) is that in the former
the correction it provides to $\alpha_0$ is of the order
of $\omega$ (an odd order contributing to the phase), and in the
latter the correction to $\beta_0$ is of the order of $\omega^2$ 
(an even
order contributing to the amplitude). Since $\alpha_0$
is of the order of $\omega$, the correction in $\alpha_\eta$ 
appears comparable to $\alpha_0$. More crucially, since $\beta_0$ 
is of the order of $\omega^0$, the correction in $\beta_\eta$ 
appears to be dominant over $\beta_0$ for large $\omega$. 
This, however, is not really the case. 
We have obtained the results given by
Eqs.~(\ref{alphacor}) and~(\ref{betacor}) under the restriction
that $\eta_\star \ll (\lambda/r_{\mathrm c})^2$. Considering 
the wavelength as $\lambda (r) =2\pi (v_0\mp a_0)/\omega$, 
we immediately see that
$\eta_\star \omega$ in Eq.~(\ref{alphacor}) and
$\eta_\star \omega^2$ in Eq.~(\ref{betacor}),
reduce both $\alpha_\eta$ and $\beta_\eta$ to be sub-leading
to $\alpha_0$ and $\beta_0$, respectively. 
This is true over most of the spatial range of the flow, except
in the close neighbourhood of the acoustic horizon, where
$\vert v_0 \vert =a_0$. In this region, for outgoing waves against the 
inflow, $\beta_\eta$ will diverge, as Eq.~(\ref{betacor}) shows, 
while $\beta_0$ itself remains finite.  

For the travelling wave, 
$f^\prime (r,t) = e^{-\beta}\exp (i\alpha -i\omega t)$, 
from which, by extracting only the amplitude part, we get 
$\vert f^\prime (r,t) \vert \sim e^{-\beta}$. Our primary 
concern is the stability of waves propagating outwards against 
the steady~\citet{bon52} transonic inflow, for which $v_0 <0$. 
We write $v_0 = -\vert v_0 \vert$ for the~\citet{bon52} inflow, 
and choose the lower sign in Eq.~(\ref{betacor}) for outgoing waves. 
These specifications give   
\begin{equation}
\label{effplus} 
\vert f^\prime (r,t) \vert \sim \frac{1}{\sqrt{a_0 \vert v_0 \vert}}
\exp \left[
\int 
\frac{\eta \omega^2 \vert v_0 \vert}{2\rho_0 a_0
\left(\vert v_0 \vert -a_0\right)^3}\,
{\mathrm d}r \right].
\end{equation}
In the supersonic region, bounded within the spherical sonic horizon, 
$\vert v_0 \vert > a_0$. For a wave approaching the sonic horizon from 
the supercritical side, with $\vert v_0 \vert \longrightarrow a_0$, 
the integrand in Eq.~(\ref{effplus}) suffers a divergence, 
and $\vert f^\prime (r,t) \vert \longrightarrow \infty$. 
The instability of the wave amplitude stands out very clearly in 
this case.  
In contrast, in the subsonic region just outside the sonic 
horizon, where $\vert v_0 \vert < a_0$, the same integrand
acquires a negative sign overall, and 
$\vert f^\prime (r,t) \vert \longrightarrow 0$. 
With a small viscosity, the sonic horizon of~\citet{bon52} accretion 
forces a discontinuity in the outward 
propagation of the wave, and acts like an impenetrable barrier 
to block acoustic waves within itself.\footnote{
A viscosity-induced divergence of the wave amplitude near an 
acoustic horizon, as Eq.~(\ref{effplus}) indicates, is similarly 
seen in the two-dimensional outflow of the shallow-water hydraulic 
jump, where a perturbation in the sub-critical region, propagating 
against the steady outflow, accumulates a wall of water at the 
horizon~\citep{rb07,kdc07}. However, 
the propagating perturbation does not breach
the horizon, thus characterizing it as a white hole~\citep{rb07,bhat17}. 
What is more, specific to astrophysical accretion itself, 
a small viscosity can cause a perturbation to diverge, leading 
to a secular instability~\citep{br07}.}   

The energy flux of the perturbation also behaves in a manner 
similar to its amplitude. The kinetic energy per unit volume 
is $E_\mathrm{kin}= (1/2)(\rho_0 +\rho^\prime)
(v_0 + v^\prime)^2$~\citep{pso80}. The potential energy per 
unit volume, with contributions from both the gravitational 
energy and the internal energy~\citep{pso80}, is 
$E_\mathrm{pot}= (\rho_0 +\rho^\prime)
(GM/r)+\rho_0 \epsilon + 
[\partial (\rho_0 \epsilon)/\partial \rho_0] \rho^\prime +(1/2)
[\partial^2 (\rho_0 \epsilon)/\partial \rho_0^2]{\rho^\prime}^2$,
with $\epsilon$ being the internal energy per unit mass~\citep{ll87}.
In both of the foregoing expressions of energy, the zeroth-order
terms refer to the steady flow, and the first-order terms
disappear upon time-averaging.
Thereafter, the time-averaged total energy in the perturbation,
per unit volume of fluid,
is to be obtained by summing the second-order terms in
$E_\mathrm{kin}$ and $E_\mathrm{pot}$.
All of these terms go either as ${\rho^\prime}^2$ or
${v^\prime}^2$, or as a product of $\rho^\prime$ and $v^\prime$. 
Making use of
Eqs.~(\ref{pertef}),~(\ref{pertrho}) and~(\ref{alphanot}), we get
$(\rho^\prime/\rho_0) \simeq v_0(v_0 \mp a_0)^{-1}
(f^\prime/f_0)$ and 
$(v^\prime/v_0) \simeq \mp a_0(v_0 \mp a_0)^{-1} 
(f^\prime/f_0)$.
Once we have two relations explicitly connecting $\rho^\prime$
and $v^\prime$ with $f^\prime$, 
the time-averaged total energy per unit volume is 
$E_\mathrm{tot} \sim \langle {\vert f^\prime (r,t) \vert}^2\rangle$. 
The energy flux of the spherical wavefront, moving with the 
speed, $(v_0 \mp a_0)$, is $F=4\pi r^2 E_\mathrm{tot}(v_0\mp a_0)$. 
Clearly, for the wave travelling outwards against
the~\citet{bon52} inflow, very close to the sonic horizon, 
both $E_\mathrm{tot}$ and $F$ will exhibit the same instability 
implied by Eq.~(\ref{effplus}).  

\section{Residues and acoustic tunnelling through the horizon} 
\label{sec6}
Waves propagating outwards against the steady~\citet{bon52} 
inflow (for which $v_0 =-\vert v_0 \vert$),
encounter a singularity at the sonic horizon, where 
$\vert v_0 \vert =a_0$. This is obvious from the integrands in 
Eqs.~(\ref{alphanot}),~(\ref{alphacor}) and~(\ref{betacor}). 
In each case, circumvention of the singularity requires 
rendering it as a simple pole on the path of the integration, 
and then applying Cauchy's residue theorem on the path.
We first demonstrate this procedure in full for the simplest of the 
cases, which is in Eq.~(\ref{alphanot}), by considering its lower 
sign, as only this pertains to an outward wave against 
the inflow. The main contribution 
to the integral comes from the immediate neighbourhood of 
$\vert v_0 \vert =a_0$, which is also where $r=r_{\mathrm c}$,
as Eqs.~(\ref{vrcrit}) show. 
A Taylor expansion about the horizon, up to the first order, gives 
$a_0 -\vert v_0 \vert \simeq (a_0 -\vert v_0 \vert)_{r_{\mathrm c}}
+[{\mathrm d}(a_0 -\vert v_0 \vert)/{\mathrm d}r]_{r_{\mathrm c}}
(r-r_{\mathrm c})$. The Taylor expansion in the neighbourhood of 
the horizon transforms the singularity at $\vert v_0 \vert =a_0$ 
to a simple pole at $r=r_{\mathrm c}$. Going by what 
Eqs.~(\ref{vrcrit}) suggest, the zero-order term in the Taylor 
expansion vanishes, in consequence of which, we approximate 
Eq.~(\ref{alphanot}) as 
\begin{equation}
\label{alphahor} 
\alpha_0 \simeq \int \frac{\omega}{
\left[{\mathrm d}(a_0-\vert v_0\vert)/{\mathrm d}r\right]_{r_{\mathrm c}}
\left(r-r_{\mathrm c}\right)}\,{\mathrm d}r.
\end{equation}
The analogue surface gravity,  
$g_\mathrm{s}=
a_\mathrm{c}[{\mathrm d}(a_0 -\vert v_0\vert)/
{\mathrm d}r]_{r_{\mathrm c}}$, at the sonic horizon~\citep{vis98}, 
and the analogue Hawking temperature,  
$T_\mathrm{H}=(\hbar g_\mathrm{s})/(2\pi k_\mathrm{B} 
a_\mathrm{c})$~\citep{vis98}. In terms of $g_\mathrm{s}$ and 
$T_\mathrm{H}$, the integral in Eq.~(\ref{alphahor}), 
taking the residue at the pole, becomes 
\begin{equation} 
\label{alphres0} 
\alpha_0 \simeq 
\frac{2\hbar \omega a_\mathrm{c}}{2\hbar g_\mathrm{s}} 
\left(\pm i \pi \right)+{\mathcal P}\left[\alpha_0\right]
=\frac{\hbar \omega}{2k_\mathrm{B}T_\mathrm{H}} 
\left(\pm i\right)+{\mathcal P}\left[\alpha_0\right], 
\end{equation} 
where ${\mathcal P}[\alpha_0]$ is the 
principal value of the integral. 
Furthermore, the negative sign in 
$\pm i$ is due to a clockwise detour of 
the pole, and the positive sign is due to an anti-clockwise
detour. Both are valid mathematically, but 
in a real physical sense, the ultimate choice of the sign 
is determined by the boundary condition at the 
pole~\citep{dk96}. Since $\alpha =\alpha_0+\alpha_\eta$, 
the imaginary part of $\alpha_0$ in Eq.~(\ref{alphres0})
contributes to the amplitude of 
$f^\prime (r,t) = e^{-\beta}\exp (i\alpha -i\omega t)$. 
Now, 
the horizon
acts like an unyielding barrier to outgoing waves. This boundary 
condition at the horizon necessitates the choice of the positive
sign of $i$ in Eq.~(\ref{alphres0}), and as such, the wave, 
which can only be very weak 
with a decaying amplitude, tunnels through the barrier.  

The tunnelling amplitude receives an additional contribution 
from $\alpha_\eta$, as it is given in Eq.~(\ref{alphacor}), with 
the integral being a sum of two terms. For simplicity of notation, 
we write $\gamma =2+(4r/v_0){\mathrm d}v_0/{\mathrm d}r$. 
When the inflow is in free fall close to the accretor, 
$v_0 \sim r^{-1/2}$, and when the inflow is highly subsonic far
away from the accretor, $v_0 \sim r^{-2}$~\citep{pso80}. In 
the former case, $\gamma =0$, and in the latter, $\gamma =-6$. 
We expect $\gamma (r_\mathrm{c})$ to have an intermediate 
value between these two limits. Following what we did to 
arrive at Eq.~(\ref{alphahor}), a Taylor expansion up to 
the first order gives 
$(a_0 -\vert v_0 \vert)^2 \simeq 
(a_0 -\vert v_0 \vert)_{r_{\mathrm c}}^2 
+2(a_0-\vert v_0 \vert)_{r_{\mathrm c}}
[{\mathrm d}(a_0 -\vert v_0 \vert)/{\mathrm d}r]_{r_{\mathrm c}}
(r-r_{\mathrm c})$. Now that we explicitly account for a 
small order of viscosity, we see from Eq.~(\ref{vcrit}) 
that $(a_0 -\vert v_0 \vert)_{r_{\mathrm c}}^2$ is 
a small non-vanishing 
quantity of the order of $\eta_\star^2$. Set against this, the 
first-order term in the Taylor expansion is of the order of 
$\eta_\star$. This argument allows us to neglect the zero-order
term in the Taylor series, and write 
$(a_0 -\vert v_0 \vert)^2 \simeq 
-2(\vert v_0 \vert -a_0)_{r_{\mathrm c}}
[{\mathrm d}(a_0 -\vert v_0 \vert)/{\mathrm d}r]_{r_{\mathrm c}}
(r-r_{\mathrm c})$. By the same token, we also approximate
$(a_0 -\vert v_0 \vert)^3 \simeq 
3(a_0-\vert v_0 \vert)_{r_{\mathrm c}}^2
[{\mathrm d}(a_0 -\vert v_0 \vert)/{\mathrm d}r]_{r_{\mathrm c}}
(r-r_{\mathrm c})$.
These conditions, imposed about the sonic horizon (where
the singularity contributes the most to the integral),  
approximates Eq.~(\ref{alphacor}), with its lower sign, to 
\begin{multline} 
\label{alphacorhor} 
\alpha_\eta \simeq  
\int \frac{\eta \omega \vert \gamma (r_\mathrm{c})\vert}
{4\rho_\mathrm{c} r_\mathrm{c}(\vert v_0 \vert -a_0)_{r_{\mathrm c}} 
\left[{\mathrm d}(a_0-\vert v_0\vert)/{\mathrm d}r\right]_{r_{\mathrm c}}
\left(r-r_{\mathrm c}\right)}\,{\mathrm d}r \\
-\int
\frac{\eta \omega}
{2\rho_\mathrm{c}(a_0-\vert v_0 \vert)_{r_{\mathrm c}}^2 
\left(r-r_{\mathrm c}\right)}\,{\mathrm d}r. 
\end{multline}
About the sonic horizon, the kinematic viscosity,
$\nu_\mathrm{c}=\eta/\rho_\mathrm{c}$. The wave number, 
$k(r)=\omega/(\vert v_0\vert -a_0)$, which is  
blue-shifted near the horizon~\citep{pw2000}.
We define a frequency,  
$\Omega
=\nu_\mathrm{c} \vert k(r_\mathrm{c})\vert/r_\mathrm{c}$,  
and a temperature, 
$T_\mathrm{s}=[\hbar (a_0-\vert v_0 \vert)_{r_{\mathrm c}}]/ 
(2\pi k_\mathrm{B} r_\mathrm{c})$. With these definitions, and by
taking the residue at the pole, the integral in Eq.~(\ref{alphacorhor}) 
gives 
\begin{equation}
\label{alphreseta} 
\alpha_\eta \simeq \pm i
\left[\frac{\hbar \Omega \vert \gamma (r_\mathrm{c})\vert}
{8k_\mathrm{B} T_\mathrm{H}}+ \frac{\hbar \Omega}
{4k_\mathrm{B} T_\mathrm{s}}\right] + 
{\mathcal P}\left[\alpha_\eta \right],
\end{equation} 
with ${\mathcal P}[\alpha_\eta ]$ being the principal
value of the integral. The singularity, as noted earlier, stands 
as a barrier against 
an outgoing acoustic wave that has to pass from the supersonic 
region to the subsonic region. This being the physical boundary 
condition at the singularity, a wave can at best tunnel through 
it with a decaying amplitude. 
Since $\alpha_\eta$ appears in the 
phase of $f^\prime (r,t) \sim \exp (i\alpha -i\omega t)$ through 
$\alpha = \alpha_0 + \alpha_\eta$, 
to make for a decaying amplitude, we select the positive sign of 
$\pm i$ in Eq.~(\ref{alphreseta}). Likewise, not to forget what 
comes with the positive sign of $\pm i$ in Eq.~(\ref{alphres0}), 
the total contribution to the amplitude of the tunnelling wave,
from both Eqs.~(\ref{alphres0}) and~(\ref{alphreseta}), is 
written as $\vert f_\mathrm{T}^\prime \vert \sim e^{-\Gamma}$, 
in which 
\begin{equation} 
\label{amplitun} 
\Gamma = 
\frac{\hbar \omega}{2k_\mathrm{B}T_\mathrm{H}} +
\frac{\hbar \Omega \vert \gamma (r_\mathrm{c})\vert}
{8k_\mathrm{B} T_\mathrm{H}}+ \frac{\hbar \Omega}
{4k_\mathrm{B} T_\mathrm{s}}. 
\end{equation} 
With $\vert f_\mathrm{T}^\prime \vert$ determined thus, the 
tunnelling probability is found from  
$\vert f_\mathrm{T}^\prime \vert^2$. The second term 
in Eq.~(\ref{amplitun}) is of particular interest to us. 
The frequency, $\Omega$, is dependent on viscosity, and it 
is the dissipative influence of viscosity that also shrinks 
the sonic sphere, as we deduce from Eq.~(\ref{rcrit}). 
What is more, in the tunnelling amplitude, $\hbar \Omega$ 
is scaled by the fluid analogue 
of the Hawking temperature. The combined effect of these facts 
is that the second term in Eq.~(\ref{amplitun}) is  
responsible for the phenomenon of black hole evaporation by phonon 
radiation in fluid analogues~\citep{wgu81,tj91,wgu95}. 
Especially when, with a large blue-shift near the horizon, 
the corresponding increase in momentum can cause a strong 
dissipation~\citep{rp2015}. 
However, we also note 
that $T_\mathrm{s} \ll T_\mathrm{H}$, from which we realize
that the third term in Eq.~(\ref{amplitun}) will overwhelm
both the terms scaled by $T_\mathrm{H}$. We 
consider the physical meaning of $T_\mathrm{s}$ vis-\`a-vis 
$T_\mathrm{H}$. Phonons are physically valid only as long as 
the acoustic structure of the fluid flow is preserved.
The thermal flux of these phonons through the acoustic horizon
has the temperature, $T_\mathrm{H}$~\citep{wgu81,tj91}. 
Close to the horizon, the outgoing waves 
experience a blue-shift~\citep{pw2000}, with 
$(a_0-\vert v_0 \vert)_{r_{\mathrm c}} \longrightarrow 0$. 
The high momentum (and the short wavelength) of the 
blue-shift disrupts the acoustic structure~\citep{vis98}. 
Physically this happens because the wavelength of the 
acoustic wave becomes small enough to be comparable to the 
mean free path of the molecules in the fluid~\citep{vis98}. 
If we go by the definition of $T_\mathrm{s}$, we see 
that with the blue-shift near the acoustic horizon, 
$T_\mathrm{s} \longrightarrow 0$.  
Since the same blue-shift violates the 
acoustic structure, $T_\mathrm{s}$ 
should not be associated with the thermal phonon flux. It is 
named ``temperature" because of its dimensional compatibility 
with $T_\mathrm{H}$. One might conjecture that $T_\mathrm{s}$ 
represents the thermal state of the phonon flux when the 
acoustic geometry is on the verge of breaking down. 

From the perspective of the acoustic continuum, 
the blue-shift itself calls for careful consideration. 
For the radially convergent flow of an inviscid compressible 
fluid, 
the sonic 
condition at the horizon is $\vert v_0 \vert =a_0$, because 
of which, for a wave propagating outwards against the flow, 
the group velocity, $v_{\mathrm g} \longrightarrow 0$. This
much is clear from Eq.~(\ref{groupvel}) when $\eta =0$. 
The wavenumber, 
$k=\omega/(\vert v_0\vert -a_0)$, is then blue-shifted without
limit near the horizon, and the corresponding wavelength is 
shortened arbitrarily. As long as the acoustic continuum holds,
the blue-shifting can continue, but physically it is limited 
by the molecular nature of matter. On the molecular scale,  
arbitrarily short wavelengths will eventually become comparable 
to the intermolecular spacing, and 
the fluid continuum needed for the acoustic propagation will 
no longer hold. Waves travelling outwards 
from the acoustic horizon can be traced back in time towards
the horizon itself. This will entail the blue-shifting 
of the wave number. But even before the horizon can be 
reached by the wave travelling backward, arbitrary blue-shifting 
will ensure the breakdown of the acoustic continuum. Therefore, 
with the loss of the acoustic continuum at very short 
wavelengths, the acoustic wave may never seem to have originated 
at the horizon or arbitrarily close to it. This difficulty, 
known for long, has been taken up and addressed 
variously~\citep{tj91,wgu95,bmps95,cj96}. With viscosity 
included in the fluid, however, the physical conditions become
qualitatively different. 
From Eqs.~(\ref{intervcrit}) and~(\ref{groupvel}), we realize 
that viscosity creates a thin layer of uncertainty about the 
exact transonic condition of the inviscid state. With 
$\vert v_0 \vert -a_0$ restricted to a small non-zero value by 
viscosity (instead of just 
vanishing at the horizon), it is possible for an outgoing 
wave packet to avoid an infinite blue-shift and emerge from 
the uncertain layer about the horizon. 
Another consequence of viscosity is that time reversal, 
$t \longrightarrow -t$, will no 
longer be symmetric, as shown by the first-order time 
derivative in Eq.~(\ref{visweq}). Hence, 
a wave packet, made to propagate backward in time, will
no longer trace back the same physical states of its forward 
propagation in time.  

Change of entropy also prevents time reversal. 
In a polytropic process, the entropy, $\mathrm S$, changes 
according to~\citep{sc39} 
\begin{equation} 
\label{dentro} 
{\mathrm d}{\mathrm S} = c_{\mathrm V}
\left[\frac{{\mathrm d}K}{K} 
+\left(\gamma -\frac{c_{\mathrm P}}{c_{\mathrm V}}\right) 
\frac{{\mathrm d}\rho}{\rho}\right].
\end{equation}
While Eq.~(\ref{dentro}) gives a general thermodynamic 
perspective, the specific mechanical means of entropy change 
in the astrophysical fluid flow of our present interest is viscosity. 
It causes heating of the fluid, and this heat is then dissipated 
to the surroundings by radiative processes. This sequence of 
events, taken together, increases the universal entropy and is 
fundamentally irreversible. As it happens, the same viscous 
dissipation breaks the time-reversal symmetry of the 
propagation of an acoustic wave group, as Eq.~(\ref{visweq}) 
indicates. Thus, the breaking of the time-reversal symmetry due 
to viscous dissipation is consistent with the change of entropy.  

The residues in the singularities of $\alpha_0$ and $\alpha_\eta$
have contributed to the amplitude of the tunnelling wave. In  
reciprocation, the residue in the singularity of $\beta_\eta$, 
as shown by Eq.~(\ref{betacor}), will contribute to the phase of 
the tunnelling wave. This is found to be 
$\beta_\eta \simeq 
[(\pm i \hbar \Omega_\mathrm{T})/(12k_\mathrm{B}T_\mathrm{H})] 
+{\mathcal P}[\beta_\eta]$,
in which the tunnelling frequency,  
$\Omega_\mathrm{T} =\nu_\mathrm{c}\vert k(r_\mathrm{c})\vert^2
\gg \Omega$, and 
${\mathcal P}[\beta_\eta]$ is the principal value of the integral
in Eq.~(\ref{betacor}).
We observe that through the singularity, the mutual 
exchange of amplitude and phase, between $\alpha$ and $\beta$, 
respectively, occurs only when $\eta \neq 0$.  

\section{Concluding remarks}
\label{sec7}
We have treated viscous spherically symmetric transonic 
accretion as an astrophysical model to demonstrate 
the analogue Hawking radiation of phonons through the sonic 
horizon of an acoustic black hole. Viscosity plays a part 
in this process, as well as in shrinking the acoustic
black hole. Both phenomena are consistent with each other, 
when we recall the possibility of the evaporation of black 
holes~\citep{wgu81,tj91,wgu95}. Since viscosity appears to 
be instrumental in the Hawking radiation, which is fundamentally
a quantum effect, we refer to a study~\citep{bbdr09} that 
shows an equivalence between viscosity in the Navier-Stokes 
equation and Planck's constant in Schr\"odinger's equation. 

In the physical flow that we have studied here, viscosity has 
a weak perturbative presence about an inviscid background. 
In astrophysical accretion, however, viscosity is enhanced
because of turbulence in the flow, without which, for instance, 
the outward transport of angular momentum in accretion discs is 
not feasible~\citep{bh98,fkr02}. In spherically symmetric accretion, 
the coupling of the mean flow and the turbulent fluctuations
scales viscosity up significantly, and shifts 
the sonic horizon inwards~\citep{rb05}, qualitatively in the
same way that we have seen here. Hence, we believe 
that an effective ``turbulent viscosity"~\citep{rb05} in transonic 
inflows, can markedly alter the tunnelling amplitude of 
acoustic waves. That said, we should 
remember that turbulence is a nonlinear phenomenon, while the 
results of our present study are based on linearization. 

Astrophysical flows often possess angular 
momentum and undergo an axially symmetric differential 
rotation, as happens in accretion discs~\citep{bh98,fkr02}.
Sub-Keplerian accretion discs with low angular momentum can be 
treated as conservative flows~\citep{ray03a}, and as such they 
are models 
of analogue gravity as well~\citep{crd06,dbd07,nard12,bsr14}. 
Complications arise for axially symmetric flows 
with high angular momentum, because in such cases the infall 
requires a physical means for the outward transport 
of the angular momentum~\citep{fkr02}. For this purpose, 
turbulence, with an enhanced effective viscosity, serves 
well~\citep{bh98,fkr02}, but with
turbulence we lose the basic condition of analogue-gravity 
modelling, namely, an inviscid and vorticity-free potential 
flow~\citep{vis98}, which, otherwise, is an easy condition to 
fulfil in the spherically symmetric geometry. 
Hence, in axially symmetric geometry, 
the modelling of analogue gravity is limited to flows of 
low angular momentum, especially when such a flow is driven 
by the powerful gravity of a black hole~\citep{dbd07}. 

The physical framework of analogue gravity, the basic subject of 
our study here, is an inviscid, irrotational and vorticity-free
fluid flow in the Newtonian construct of flat space and 
time~\citep{wgu81,tj91,wgu95,vis98}. On a closed surface in 
this fluid, where the flow becomes supersonic, the acoustic 
analogue of a black-hole horizon is formed.
This is the fluid-flow model of a black hole, and the 
propagation of acoustic waves in the supersonic fluid flow 
mimics the propagation of scalar waves in the spacetime of 
a black hole~\citep{wgu81,wgu95}. 
As long as an equation of state provides 
a means for acoustic propagation, transonic hydrodynamic flows 
can generally produce an analogue metric and an acoustic horizon. While 
this appears to be a universal feature of flows that pass through 
a well-defined critical point~\citep{ncbr07,rb07,sbbr13,sr14}, 
the symmetric form of the acoustic metric can be variously 
disrupted because of the physical circumstances of a given 
fluid flow. The coupling of the flow and the geometry of 
Schwarzschild spacetime is one example of how the acoustic 
mertic can be affected adversely~\citep{ncbr07}. 
The same behaviour is also exhibited 
due to viscous dissipation in the shallow-water circular
hydraulic jump~\citep{rb07}. 
Dispersion, arising out of interactions between baryons and 
vector mesons in a nuclear outflow, similarly breaks the     
symmetry of the acoustic metric~\citep{sbbr13}. Perturbations
of nonlinear order in spherically symmetric accretion shift 
the acoustic horizon about its static position, although the 
symmetry of the acoustic metric remains intact~\citep{sr14}. 
We surmise that some of the aforementioned fluid 
systems can help to detect the  
analogue Hawking radiation. 

\begin{acknowledgments}
This research has made use of NASA's Astrophysics Data System.
The author thanks J. K. Bhattacharjee, T. K. Das and T. Jacobson 
for their comments.  
He is also grateful for the library support of Jaypee University 
of Engineering and Technology, Raghogarh, Guna, India. 
\end{acknowledgments}

\bibliography{akr0920}
\end{document}